\documentclass[aps,prl,superscriptaddress,twocolumn]{revtex4-1}
\usepackage{subfigure,graphics} 
\usepackage{amsfonts,amssymb,amsmath}
\usepackage{flafter}
\usepackage{multirow}
\usepackage{txfonts}
\usepackage[unicode,breaklinks]{hyperref}
\newcommand{\papertitle}{Simulating seeded vacuum decay in a cold atom system}
\hypersetup{
    unicode=true,
    a4paper=true,
    plainpages=false,
    pdftitle={\papertitle},
    pdfauthor={Thomas P. Billam, Ruth Gregory, Florent Michel, Ian G. Moss},
    pdfsubject={\papertitle},
    colorlinks=true,
    linkcolor=blue,
    citecolor=blue,
    filecolor=black,
    urlcolor=blue
}

\usepackage{ulem}

\newcommand{\be}{\begin{equation}}
\newcommand{\ee}{\end{equation}}
\newcommand{\bea}{\begin{eqnarray}}
\newcommand{\eea}{\end{eqnarray}}
\newcommand{\beal}{\begin{aligned}}
\newcommand{\eeal}{\end{aligned}}

\begin{document} 

\title{\papertitle}

\author{Thomas P.\ Billam}
\email{thomas.billam@ncl.ac.uk}
\affiliation{Joint Quantum Centre (JQC) Durham--Newcastle, School of Mathematics, Statistics and Physics, 
Newcastle University, Newcastle upon Tyne, NE1 7RU, UK}

\author{Ruth Gregory}
\email{r.a.w.gregory@durham.ac.uk}
\affiliation{Centre for Particle Theory, Durham University, South Road, 
Durham, DH1 3LE, UK}
\affiliation{Perimeter Institute, 31 Caroline Street North, Waterloo, 
ON, N2L 2Y5, Canada}

\author{Florent Michel}
\email{florent.c.michel@durham.ac.uk}
\affiliation{Centre for Particle Theory, Durham University, South Road, 
Durham, DH1 3LE, UK}

\author{Ian G. Moss}
\email{ian.moss@ncl.ac.uk}
\affiliation{School of Mathematics, Statistics and Physics, 
Newcastle University, Newcastle upon Tyne, NE1 7RU, UK}

\date{\today}

\begin{abstract}
We propose to test the concept of seeded vacuum decay in cosmology using a
Bose-Einstein condensate system. The role of the nucleation seed is played by a
vortex within the condensate. We present two complementary theoretical
analyses that demonstrate seeded decay is the dominant decay mechanism of the
false vacuum.  First, we adapt the standard instanton methods to the
Gross-Pitaevskii equation. Second, we use the truncated Wigner method to study
vacuum decay. 
\end{abstract}
\pacs{PACS number(s): }

\maketitle

First-order phase transitions form an important class of physical phenomena.
Typically, these are characterised by metastable, supercooled states and the nucleation of bubbles. 
Applications range from the condensation of water vapour to the vacuum decay of fundamental quantum
fields. In cosmology, bubbles of a new matter phase would produce huge density variations, and 
unsurprisingly first order phase transitions have been proposed as sources of gravitational 
waves
\cite{Caprini:2009fx, Hindmarsh:2013xza}
and as sources of primordial black holes 
\cite{Hawking:1982ga, Deng:2017uwc}.

Clearly a key factor in the relevance of such by-products of phase transitions
is the likelihood of that transition occurring. Bubble nucleation rates are exponentially suppressed,
and formal estimates of the lifetimes of metastable states can be huge. However, many phase transition 
rates in ordinary matter are greatly enhanced by the presence of nucleation seeds, 
in the form of impurities or defects on the boundary of the material. We have 
argued recently that cosmological bubble nucleation can also be greatly accelerated by 
nucleation seeds, for example with seeds in the form of primordial black holes 
\cite{Gregory:2013hja,Burda:2015isa}.
In this paper we propose that seeded bubble nucleation can be studied in
a laboratory cold-atom analogue of cosmological vacuum decay
\cite{0295-5075-110-5-56001,0953-4075-50-2-024003}.

The idea of using analogue systems for cosmological processes comes under the 
general area of modelling the ``universe in the laboratory''~\cite{RogerNatCom16,Eckel:2017uqx}.
So far, analogue systems have mostly been employed to test ideas in
perturbative quantum field theory~\cite{Unruh:1980cg, Barcelo:2005fc}, 
but nonperturbative phenomena such as
bubble nucleation also play an important role in quantum mechanics and field theory.

As pointed out in the classic work of Coleman and others~\cite{Coleman:1977py,
Callan:1977pt, Coleman:1980aw}, the bubble nucleation process
in quantum field theory can be described  described by an {\it instanton}, or \textit{bounce}, solution to the
field equations in imaginary time. The probability for decay is then given, to
leading order, by a negative exponential of the action of the instanton.  
Understanding vacuum decay and the role of the instanton is now particularly
pressing in light of the measurements of the Higgs mass, that currently
indicates our vacuum is in a region of metastability~\cite{Degrassi:2012ry}.

\begin{center}
\begin{figure}[htb]
\includegraphics{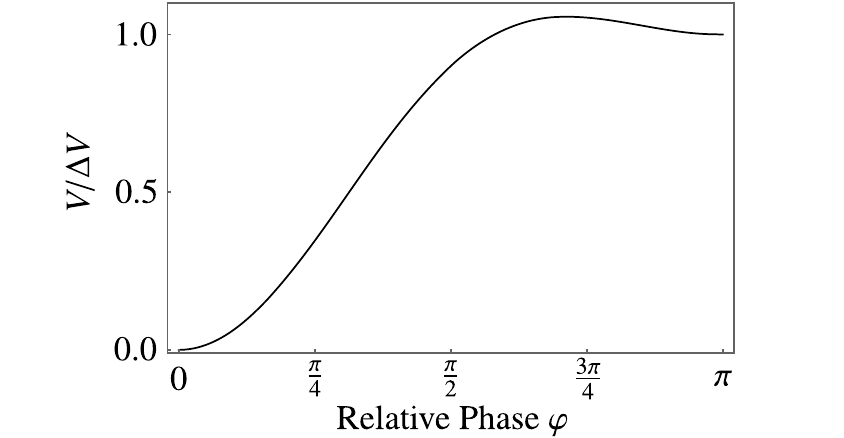}
\caption{The field potential $V$ plotted as a function of the
relative phase of the two atomic wave functions, $\varphi$. The
false vacuum is the minimum at $\varphi=\pi$ and the true
vacuum the global minimum at $\varphi=0$. $\Delta V$ is the difference
in vacuum energy.  } \label{pot} 
\end{figure} 
\end{center}

The semi-classical description of vacuum decay with gravity involves
analytically continuing to imaginary time, and finding the gravitational
instanton. However, while most are comfortable with the assumptions used in
perturbative quantum field theory on curved spacetime, such non-perturbative
processes are sometimes viewed with more caution. The ability to test such a
process via an analogue ``table-top'' quantum system would be a strong
vindication of the use of such techniques.  To this end, there have been some
recent developments in exploring possible analogue systems that could test
vacuum decay.  Fialko et al \cite{0295-5075-110-5-56001,0953-4075-50-2-024003}
proposed an experiment in a laboratory cold atom system. Their system consists
of a Bose gas with two different spin states of the same atom species in an
optical trap. The two states are coupled by a microwave field.  By modulating
the amplitude of the microwave field, a new quartic interaction between the two
states is induced in the time-averaged theory which creates a non-trivial
ground state structure as illustrated in figure \ref{pot} \footnote{It has been
pointed out that there is an instability in the equations describing
the modulated system
\cite{Braden:2018tky,Braden:2017add}.
However, for the parameters used in this paper and trap radius $25\mu{\rm m}$
with ${}^{39}K$, the instability occurs for a 
modulation frequency $\omega< 800 k^2{\rm Hz}$, where $k$ is the maximum
wavenumber consistent with the equations, in units of the healing length. 
We will assume that the modulation frequency is above this limit.
}.

In this paper we propose an analogue system that can explore the process of
catalysis of vacuum decay that is central to our previous results.  We use the
above model  to test seeded vacuum decay by introducing a vortex into the two
dimensional spinor Bose gas system.  We have used two complementary theoretical
approaches. Firstly, we have applied Coleman's non-perturbative theory of
vacuum decay to the Gross-Pitaevskii equation (GPE). Secondly, we have used the
truncated Wigner method, a stochastic approach, to study the vacuum decay.
In both cases, we find that the introduction of the vortex seed enhances the
probability of vacuum decay.

Our system is a two-component BEC of atoms with mass $m$ 
coupled by a modulated microwave field.  
The Hamiltonian operator in $n$ dimensions is given by
\begin{equation}
\hat H=\int d^nx\left\{
\psi_i^\dagger\left[{-\hbar^2\nabla^2\over 2m}\right]\psi_i
+V(\psi_i,\psi_i^\dagger)\right\},
\label{hamil}
\end{equation}
with field operators  $\psi_i$, $i=1,2$ and summation over the spin indices implied.
Fialko et al. \cite{0295-5075-110-5-56001,0953-4075-50-2-024003} 
described a procedure whereby averaging 
over timescales longer than the modulation timescale
leads to an interaction potential of the form
\begin{equation}
V=\frac{g}2(\psi_i^\dagger)^2(\psi_i)^2 - \mu \psi^\dagger_i \psi_i
-\nu\psi^\dagger_i\sigma_{\!\!x\,ij} \psi_j+
\frac{g\nu\lambda^2}{4\mu}(\psi^\dagger_i\sigma_{\!\!y\,ij} \psi_j)^2,
\end{equation}
where the $\sigma_i$ are the Pauli matrices.
The potential includes the chemical potential $\mu$, 
point-like interactions of strength $g$ between the field operators
and the microwave induced interaction $\nu$. The final term comes from the 
averaging procedure and introduces a new parameter $\lambda$, dependent
on the amplitude of the modulation.
The trapping potential used to confine the condensate has been omitted in 
order to isolate the physics of vacuum decay.

The terms proportional to $\nu$ are responsible for the difference in energy 
between the global and local minima of the energy. The global minimum 
represents the true vacuum state and the local minimum represents the false 
vacuum. In order to parameterise the difference in energy between the vacua, 
we introduce a `small' dimensionless parameter $\epsilon$ by 
\begin{equation}
\epsilon=\left({\nu\over \mu}\right)^{1/2}.
\end{equation}
For $\nu>0$, the true vacuum is a state with $\psi_1=\psi_2$ and the
false vacuum is a state with $\psi_1=-\psi_2$. The condensate densities 
of the two components at the extrema are equal to one another, and 
given by $\langle \psi_1^\dagger\psi_1\rangle =\langle\psi_2^\dagger\psi_2\rangle 
=\rho_m(1\pm\epsilon^2)$. Note that we prefer to work with the mean 
density $\rho_m=\mu/g$ rather than the chemical potential.
The difference in energy density between the two vacuum states is given by 
$\Delta V=4g\rho_m^2\epsilon^2$.

The non-perturbative theory of vacuum decay starts with the imaginary-time 
partition function
\begin{equation}
Z=\int D\psi_i D\bar\psi_i\,e^{-S[\psi_i,\bar\psi_i]/\hbar},
\end{equation}
where the integral extends over complex fields $\psi_i$ and their 
complex conjugates $\bar\psi_i$ with action
\be
S[\psi_i,\bar\psi_i]=\!\!
\int\! d^nx d\tau\left\{\hbar\bar\psi_i\partial_\tau\psi_i
-\bar\psi_i{\hbar^2\over 2m}\nabla^2\psi_i
+V(\psi_i,\bar\psi_i)\right\}
\ee
Vacuum decay is associated with instanton solutions to field equations 
in imaginary time $\tau=it$ \cite{Coleman:1977py,Callan:1977pt}
\be
\beal
{\hbar^2\over 2m}\nabla^2\psi_i-\hbar\partial_\tau\psi_i
-{\partial V\over\partial\bar\psi_i} &=0, \\
{\hbar^2\over 2m}\nabla^2\bar\psi_i+\hbar\partial_\tau{\bar\psi_i}
-{\partial V\over\partial\psi_i}&=0,
\eeal
\label{bubs}
\ee
and fields that approach the false vacuum as $r,\tau\to\infty$.

On the original path integration contour, $\psi_i$ and $\bar\psi_i$ are 
complex conjugates and the field equations imply that the saddle points 
are static. In order to find the non-static bubble solutions, we have to 
deform the path of integration into a wider region of complex 
function space where  $\bar\psi_i$ is not the complex conjugate of $\psi_i$. 
Although this may appear a strange procedure at first sight, this analytic continuation 
is already implicit in the previous work on vacuum decay as we shall see later.

The full expression for the nucleation rate of vacuum bubbles in a volume ${\cal V}$ is 
\cite{Coleman:1977py,Callan:1977pt},
\begin{equation}
\Gamma \approx {\cal V}\left|
{{\rm det}'\,S''[\psi_b]\over {\rm det}\,S''[\psi_{\rm fv}]}
\right|^{-1/2}\,\left({S[\psi_b]\over 2\pi\hbar}\right)^{N/2}\,
e^{-S[\psi_b]/\hbar}.\label{gamma}
\end{equation}
where $S''$ denotes the second functional derivative of the action $S$, 
and det$'$ denotes omission of $N=n+1$ zero modes from the functional 
determinant of the operator. (For convenience, we always include a constant 
shift to the action so that the action of the false vacuum is zero.)
For seeded nucleation, the volume factor is replaced by the number 
of nucleation seeds and the number of zero modes becomes $N=1$. 
The key feature here is the exponential suppression of the 
decay rate, and the non-perturbative treatment fails if the exponent is small. 

In vacuum decay, the key quantity determining physical aspects of decay
is the energy splitting between true and false vacua, $\Delta V$, which
is proportional to $\epsilon^2$. In our system, $\epsilon$ also determines
the magnitude of the interaction between the two scalars, and for small
$\epsilon$, most of the degrees of freedom of the system decouple,
leaving an effective field theory of the relative phases of the two condensates
as explored in \cite{0295-5075-110-5-56001,0953-4075-50-2-024003}
in one spatial dimension.

Here we are interested in seeded decay, so we consider the model
in {\it two} spatial dimensions with polar coordinates $r$ and $\theta$.
The natural size of the bubble will be determined by $R_0
=\hbar(\rho_m/ m \Delta V)^{1/2}$, and the natural timescale by
$R_0/c_s$, where the sound speed $c_s=(g\rho_m/m)^{1/2}$. 
To simplify the following analysis, we rescale our dimensionful 
coordinates accordingly, and also rescale the action:
\begin{equation}
S=\hbar\rho_m R_0^2{\hat S}
\end{equation}
\begin{figure}[htb]
\scalebox{0.6}{\includegraphics{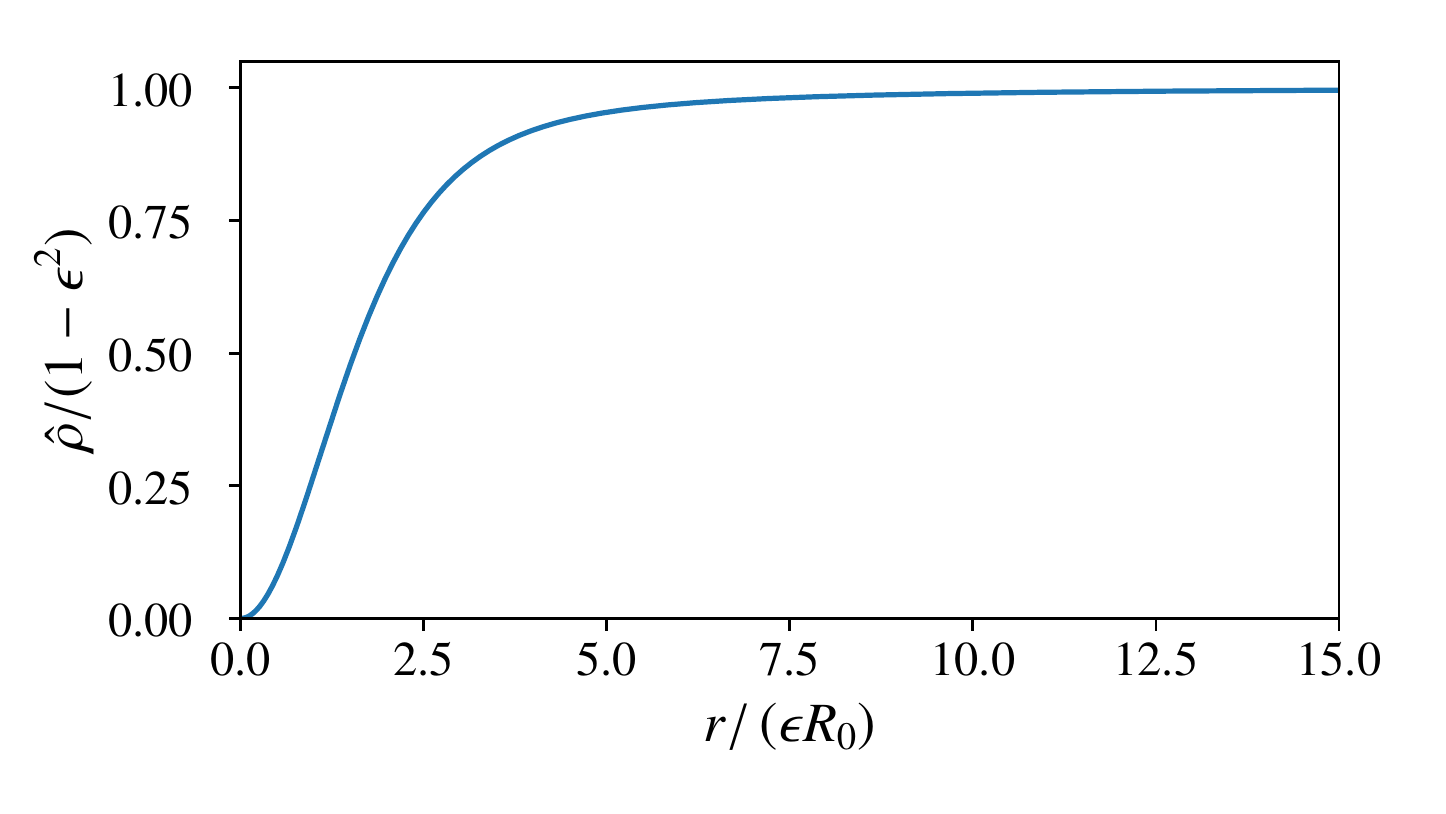}}
\caption{Vortex density profile $\hat\rho=\rho/\rho_m$ plotted as a function 
of radius $r$. The density vanishes at the centre and approaches the false 
vacuum density as $r\to\infty$. Its physical thickness scales as $\epsilon R_0$.
}
\label{profile}
\end{figure}

Since we are interested in exploring seeded decay, we look for a cylindrically
symmetric solution that explicitly highlights the relevant degrees of freedom
and includes the possibility of a topologically nontrivial vortex false vacuum state:
\be
\beal
\psi_i&=\rho^{1/2}\left(1\pm\frac\epsilon2\sigma\right)
e^{\pm i\varphi/2+in\theta},\\
\bar\psi_i&=\rho^{1/2}\left(1\pm\frac\epsilon2\sigma\right)
e^{\mp i\varphi/2-in\theta},
\eeal
\label{fields}
\ee
namely, the relative phase $\varphi$ between the two components, the 
leading order (in $\epsilon$) profile of the false vacuum background $\rho$,
an overall common phase winding $n\theta$ that is present in a nontrivial
vortex background, and the bubble profile function $\sigma$.
The upper/lower signs apply to the $i=1,2$ spin states respectively.

The pure false vacuum has $n=0$, and $\rho = \rho_m(1-\epsilon^2)$,
with instanton profiles for $\varphi$ explored in 
\cite{0295-5075-110-5-56001,0953-4075-50-2-024003}. Here we are
interested in seeded tunnelling, so we also consider the vortex background
for $n=1$, with $\rho$ satisfying the ${\cal O}(\epsilon^2)$ background
equations obtained by substituting \eqref{fields} in \eqref{bubs}.
The profile of $\rho$ is precisely that of a superfluid (or global) vortex, 
and is illustrated in figure \ref{profile}.

The potential for the instanton solutions depends only on the relative 
phase $\varphi$ and the background density $\rho$. Our rescaling of the 
length and time coordinates means that we also rescale the potential to 
$\hat V=(V-V_{TV})/2\Delta V$,
\begin{equation}
\hat V=\hat\rho(1-\cos\varphi)+\frac12\lambda^2\hat\rho^2\sin^2\varphi,
\end{equation}
as plotted in Fig. \ref{pot}. At zeroth order in $\epsilon$, the field 
equations \eqref{bubs} imply that $\sigma=-i\hat\rho^{-1}\partial_\tau\varphi$.
Note that $\sigma$ is imaginary, and the bubble solution has 
$\bar\psi_1\ne \psi_1^\dagger$ as was mentioned earlier.
Replacing $\sigma$ in the action using this field equation gives an action 
depending only on $\varphi$ which was used in 
Refs. \cite{0295-5075-110-5-56001,0953-4075-50-2-024003}.
However, at the core of the vortex, $\hat\rho\to 0$ and this 
replacement of $\sigma$ is no longer valid.
Instead, numerical solutions have been obtained by solving the full 
equations for the phase $\varphi$ and the density variation $\sigma$.

The vacuum decay rate around a single vortex, using Coleman's formula (\ref{gamma}) is
\begin{equation}
\Gamma=A{c_s\over R_0}\left({\rho_m R_0^2\hat S\over 2\pi}\right)^{1/2}
e^{-\rho_mR_0^2\hat S},
\end{equation}
where $A$ is a dimensionless numerical factor depending on the ratio of determinants
(which we do not evaluate here).
\begin{center}
\begin{figure}[htb]
\includegraphics{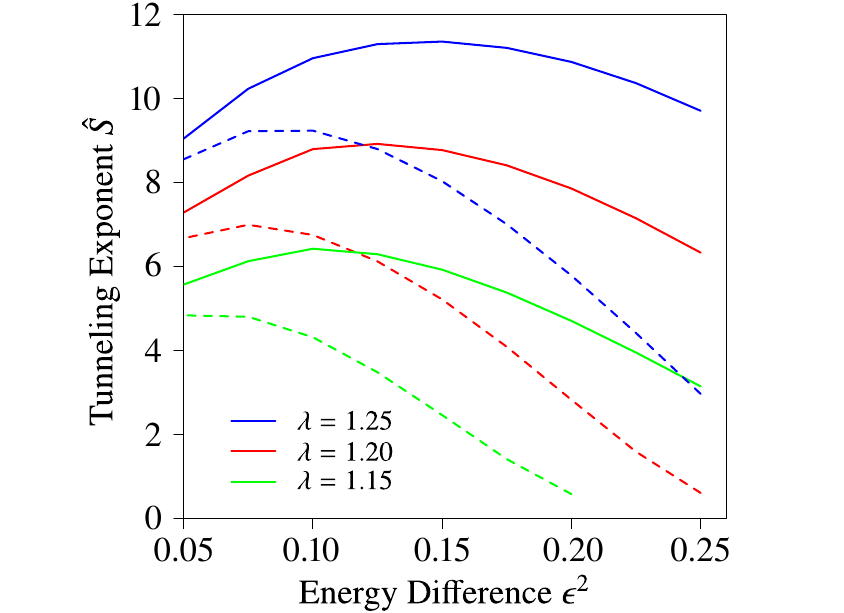}
\caption{The dimensionless exponent $\hat S$ of the vacuum decay rate 
plotted as a function of the parameter $\epsilon^2$. The solid lines represent 
unseeded vacuum decay and the dashed lines are for bubbles seeded by 
vortices. The action is lower for the seeded bubbles.
}
\label{exponent}
\end{figure}
\end{center}

Numerical results for the factor $\hat{S}$ in the decay exponent are shown in
Fig \ref{exponent}~\footnote{We have checked these results using an ansatz
approach akin to that of~\cite{Lee:2013ega}, extended to include dispersion.
The results agree within a few per cent.}. These show clearly that the tunnelling
exponent can be reduced significantly in the presence of a vortex.  The vortex
width from Fig. \ref{profile} is related to $\epsilon R_0$. Consequently,
smaller values of $\epsilon$ are associated with relatively thin vortices
compared to the bubble scale $R_0$, which have less effect on the vacuum decay
rate than vortices with larger values of $\epsilon$.

The nucleation rate depends on the physical parameters through the 
combination $\rho_m R_0^2$. The length scale $R_0$ itself is related
to the atomic scattering length $a_s$ and the thickness of the condensate 
$a_z$ via the effective coupling strength $g$, \cite{PhysRevA.65.043617},
\begin{equation}
g={4\pi \hbar^2\over m}{a_s\over \sqrt{2\pi}a_z}.
\end{equation}
Thus the factor in the decay exponent becomes 
$\rho_m R_0^2=a_z/(4\epsilon^2\sqrt{8\pi}a_s)$.

As an alternative treatment of bubble nucleation we consider a two-dimensional spinor 
BEC in a flat-bottomed optical potential. At the mean-field level,
the system can be described by the Gross-Pitaevskii equation (GPE) derived from
the symmetric Hamiltonian in the rescaled coordinates used above,
\begin{equation}
\frac{i}{\epsilon}\partial_t\psi_i=-\nabla^2\psi_i+V_T\psi_i+{\partial \hat V\over\partial\bar\psi_i}
.\label{GPE}
\end{equation}
where
\begin{equation}
{\partial \hat V\over\partial\bar\psi_i}={1\over 2\epsilon^2}\left({\bar\psi_i\psi_i\over\rho_m}-1\right)\psi_i
-\frac12(\sigma_x\psi)_i+\frac{\lambda^2}{4}{\bar\psi\sigma_y\psi\over\rho_m}(\sigma_y\psi)_i \,.
\end{equation}

The truncated Wigner approach seeks to emulate the many-body quantum field
description of a BEC with a stochastic description \cite{1402-4896-91-7-073007,
Blakie2008}. We compute the false vacuum solution to the GPE, $\psi_{i\,FV}$, 
on a square grid of side $L$ with $M$ points, creating an initial ensemble of 
fields by adding noise into unoccupied plane-wave modes according to the
zero-temperature prescription $\psi_i = \psi_{i\,FV} + f(r) \sum_{\mathbf{k}}
\beta_{i\,\mathbf{k}} e^{i \mathbf{k} \cdot \mathbf{r}}$ for all $|\mathbf{k}|
< \pi M / (2L)$, where $\beta_{i,\mathbf{k}}$ are complex gaussian random
variables with $\langle \beta_{i,\mathbf{k}}^* \beta_{j, \mathbf{k}'} \rangle =
\delta_{i,j} \delta_{\mathbf{k},\mathbf{k}'} / 2$. The function $f(r) =
\Theta(R-r)/\!\sqrt{\pi}R$ restricts the noise to the trap interior. We compute
the trajectory of each field using the GPE (using a projector to
precisely evolve the noise-seeded modes \cite{Blakie2008}), both with and
without initial imprinting of the density and phase profiles of a vortex
at the trap centre. The average $1/2$ particle
per mode of noise in each trajectory emulates vacuum fluctuations. 
\begin{figure}[ht]
\includegraphics{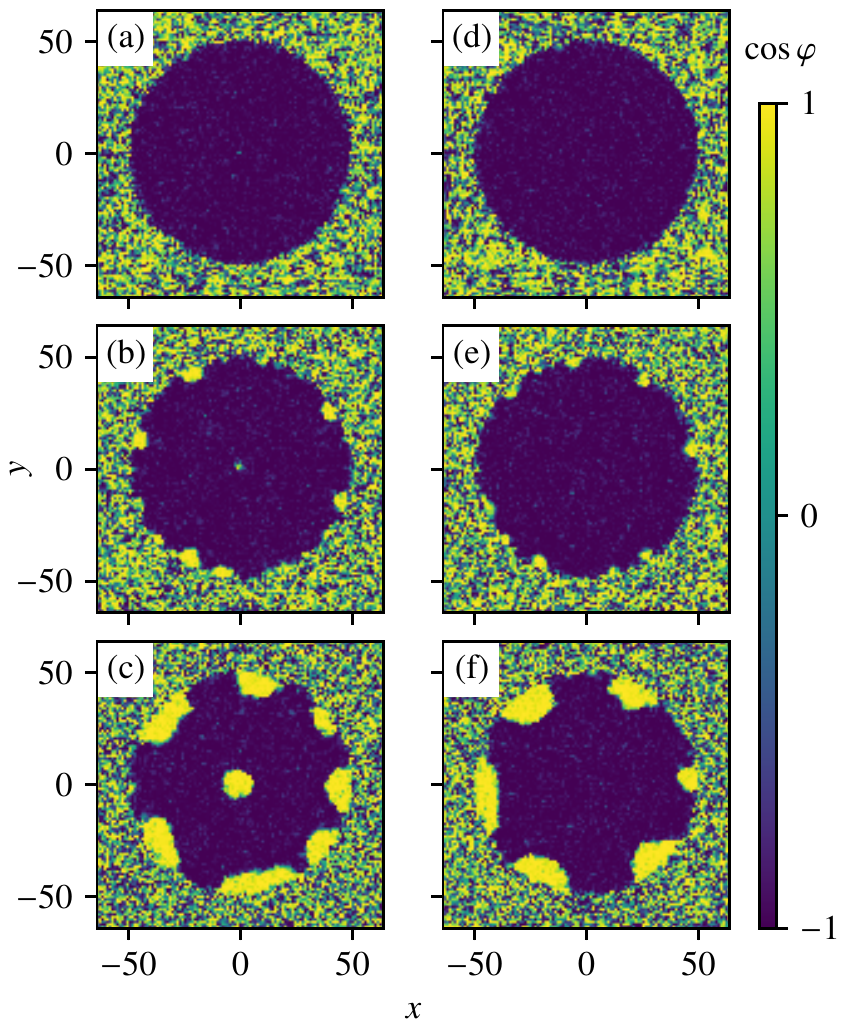}
\caption{Typical example of the decay of the false vacuum around a 
vortex in two dimensions.
The plots show the cosine of the relative phase between the two spin states.
The BEC is contained inside a circular bucket trap of radius $R = 50R_0$
[$V_T(r) = 5 (1 + \mathrm{tanh}(r-R)) / (2\epsilon)$], with dimensionless 
parameters $\lambda=1.3$, $\epsilon=0.7$ and
$\rho_m R_0^2=10.0$.  (a) Initially, the false vacuum is contained within the circular trap.  
(b) Later, a bubble of true vacuum (yellow) forms around the vortex in the centre.  
True vacuum also
forms around the walls of the trap. (c) Later still, the true vacuum regions
grow, and eventually merge. (d--f) Typical trajectory without an initially imprinted vortex
[same times as (a--c)].}
\label{twa}
\end{figure}

Numerical results, obtained partly using XMDS2 software \cite{Dennis:2013aa},
confirm that the vortex acts as a nucleation seed when the size of the vortex
is of a similar magnitude to the bubble scale $R_0$ and the particle density
$\rho R_0^2$ is not too high, just as we would expect from the non-perturbative
results in Fig.\ \ref{exponent}. In Fig.\ \ref{twa}, we demonstrate bubble
nucleation around a vortex seed (parameters in figure caption). In an ensemble
of 20 stochastic trajectories with an initially imprinted vortex evolved to
time $1500 R_0 / c_s$, the vortex seeds a bubble in 100\% of trajectories, but
a bubble nucleates spontaneously in the bulk in only 1 trajectory ($\sim$5\%).
In an ensemble of 10 stochastic trajectories with no imprinted vortex evolved
to time $2500 R_0 / c_s$, no spontaneous nucleation within the bulk was
observed.  This demonstrates the strong enhancement of bubble nucleation by the
vortex.  We also observe that the walls of the trap strongly enhance bubble
nucleation too, both with and without the imprinted vortex.  Indeed, we might
regard the region outside of the trap, where the phase oscillates widely, as
being full of `ghost' vortices. The nucleation process can be triggered by the
migration of these ghost vortices into the region just inside the wall.  Whilst
the wall effect may not be relevant to cosmology, it does introduce a new
phenomenon that will be of interest for laboratory BECs. While our simulations
represent a proof-of-principle example rather than a concrete experimental
proposal, advances in optical trapping \cite{Gaunt2013, Gauthier2016} and
various techniques for vortex imprinting in spinor condensates
\cite{Kawaguchi2012, Stamper-Kurn2013} could be used to probe similar systems
experimentally.

In conclusion, our two theoretical approaches, based on the Euclidean field
equations~\eqref{bubs} and on the truncated Wigner approximation, both show
a significant increase of the decay rate of the false vacuum in the presence of
a vortex. Numerical simulations also indicate that other kinds of
defects, such as the walls of a sharp potential trap, can have a similar
effect. Since getting a large enough decay rate is a major difficulty in
designing experiments, we expect this to be an important ingredient for putting
the theoretical model of~\cite{0295-5075-110-5-56001,0953-4075-50-2-024003} into
practice, and thus testing vacuum decay in the laboratory.

\section*{Acknowledgements} 

We would like to thank Carlo Barenghi for helpful suggestions. 
This work was supported in part by the Leverhulme Trust [grant RPG-2016-233],
the EPSRC [grant EP/R021074/1], and by the Perimeter Institute.
Research at Perimeter Institute is supported by the Government of Canada
through the Department of Innovation, Science and Economic Development
and by the Province of Ontario through the Ministry of Research and Innovation.
RG would also like to thank the Simons Foundation for support and the
Aspen Center for Physics for hospitality. 
FM thanks FAPESP for support and IFSC/USP for its hospitality.

\bibliography{paper}

\end{document}